%% file: nuK0_main.tex
\documentclass[
 reprint,
 amsmath,amssymb,
 aps,
 superscriptaddress,
]{revtex4-2}

\usepackage{graphicx}
\usepackage{dcolumn}
\usepackage{bm}
\usepackage{hyperref}
\usepackage[mathlines]{lineno}
\usepackage{multirow}

\hypersetup{
  colorlinks=true, 
  linkcolor=blue,
  citecolor=blue,
  urlcolor=blue,
  pdfborder={0 0 0}
}

\usepackage{enumerate}
\usepackage{enumitem}
\usepackage{xspace}
\usepackage{orcidlink}

\providecommand{\nuk}{\ensuremath{n\to\bar{\nu}+K^0}\xspace}
\providecommand{\nuknop}{\ensuremath{n\to\bar{\nu}K^0}\xspace}
\providecommand{\pizero}{\ensuremath{K^0_S\to2\pi^0}\xspace}
\providecommand{\pipi}{\ensuremath{K^0_S\to\pi^+\pi^-}\xspace}

\begin{document}


\title{Search for neutron decay into an antineutrino and a neutral kaon in \\
0.401 megaton-years exposure of Super-Kamiokande}

\input{authors-20250610}

\date{\today}

\begin{abstract}
We searched for bound neutron decay via $n\to\bar{\nu}+K^0$ predicted by the Grand Unified Theories in 0.401\,Mton$\cdot$years exposure of all pure water phases in the Super-Kamiokande detector.
About 4.4 times more data than in the previous search have been analyzed by a new method including a spectrum fit to kaon invariant mass distributions.
No significant data excess has been observed in the signal regions.
As a result of this analysis, we set a lower limit of $7.8\times10^{32}$\,years on the neutron lifetime at a 90\% confidence level.
\end{abstract}

\maketitle

\section{\label{sec:intro}INTRODUCTION}
Grand Unified Theories (GUTs) propose a unification of the weak, strong, and electromagnetic interactions at extremely high energies, typically the order of $10^{15}$--$10^{16}$\,GeV.
These theories predict processes that violate baryon number conservation by one unit, leading for instance to nucleon decay, which is forbidden in the Standard Model of particle physics \cite{PhysRevLett.32.438, FRITZSCH1975193}.
The baryon number violating decay of the nucleon bound inside the nucleus or the free proton decay provides a critical experimental test for these theoretical frameworks.
Some models of supersymmetric (SUSY) GUTs \cite{Hisano_1993, GOH2004105, Antusch:2020ztu} predict \nuknop as the major neutron decay channel as well as $p \to \bar{\nu} K^+$ for the proton decay.
Such a nucleon decay has not been observed. Lower limits for various channels were set by Super-Kamiokande including $p\to\bar{\nu} K^+$ \cite{PhysRevD.90.072005}.

This study reports the latest result of the search for neutron decay via \nuknop using the Super-Kamiokande (SK) water Cherenkov detector.
The SK detector is effective for searching for this decay channel because of its large fiducial mass and excellent capability to identify particle types and measure their momenta. 
This two-body decay produces an antineutrino and a neutral kaon.
The experimental signature is the decay in flight of neutral kaon with a momentum near 300\,MeV/c as seen in Figure \ref{fig:momKS}.
The recoiling neutrino is unobserved and could be either a neutrino or antineutrino.
Most GUTs conserve baryon minus lepton number (B-L) and favor an antineutrino.
However, this search is equally applicable to a neutrino in the final state, conserving B+L, as seen in some models \cite{VissaniB+L, BabuB-L}.
The neutral kaon, $K^0$, is a composite state comprising $K^0_S$ and $K^0_L$.
The $K^0_S$ meson, with a lifetime of 90\,ps, decays predominantly into $\pi^+\pi^-$ (69.2\%) and $2\pi^0$ (30.7\%).
On the other hand, $K^0_L$, with a much longer lifetime of 51\,ns, decays predominantly into $\pi^\pm l^\mp \nu$ ($l=e$ or $\mu$, 67.6\%), $3\pi^0$ (19.5\%) and $\pi^+ \pi^- \pi^0$ (12.5\%).

Search for neutron decay through the \nuknop channel has been previously conducted at SK \cite{PhysRevD.72.052007}.
The prior analysis used SK-I data, corresponding to an exposure of 92\,kton$\cdot$years collected between 1996 and 2001, and observed no significant signal excess above background for \nuknop decay.
Consequently, a lower limit on the partial lifetime of neutron for this decay channel was set as $1.3\times 10^{32}$\,years at the 90\% confidence level (C.L.).
In this study, the search was extended by a new analysis method using a complete dataset of pure water phases obtained from 1996 to 2020, which corresponds to 401\,kton$\cdot$years, 4.4 times the exposure of the previous search.

This paper is structured as follows:
Section \ref{sec:sk} describes the Super-Kamiokande detector.
Section \ref{sec:simulation} outlines the simulation methods used.
Section \ref{sec:rec} discusses improvements in event reconstruction.
The criteria for event selection and the corresponding results, including the lifetime limit, are presented in Sections \ref{sec:selection} and \ref{sec:results}, respectively.
Finally, the conclusion is summarized in Section \ref{sec:conclusion}.

\section{\label{sec:sk}SUPER-KAMIOKANDE}
Super-Kamiokande is a large water Cherenkov detector located approximately 1,000\,m underground (2,700\,m water equivalent) in the Kamioka mine, Gifu, Japan.
A cylindrical water tank with a diameter of 39.3\,m and a height of 41.4\,m holds 50\,kton of ultrapure water and detects Cherenkov light produced by charged particles using photomultiplier tubes (PMTs) arranged on its walls.
The SK detector consists of two volumes: a 32\,kton inner detector (ID) with a diameter of 33.8\,m and a height of 36.2\,m, and an outer detector (OD) surrounding the ID with $~2$\,m thickness layer.
The ID is mainly responsible for observing charged particles from nucleon decay or neutrino interaction with inward-facing 50-cm PMTs.
The main purpose of the OD is to identify cosmic ray muons entering the ID with outward-facing 20-cm PMTs.

The data used in this analysis are categorized into five distinct periods: SK-I (1996–2001), SK-II (2002–2005), SK-III (2006–2008), SK-IV (2008–2018), and SK-V (2019-2020).
The livetime is 6511.3 days in total corresponding to 0.401\,Mton$\cdot$years with a 22.5\,kton fiducial mass.
During SK-I, SK-III, SK-IV, and SK-V, more than 11,000 PMTs were placed on the ID wall with a photocathode coverage of 40\%, while it was 19\% during the SK-II period.
From SK-IV onward, the implementation of upgraded electronics enhanced the tagging efficiency of Michel electrons.
In this study, the dataset was analyzed independently of the previous analysis \cite{PhysRevD.72.052007} that was performed only for the SK-I period.
The SK detector has been comprehensively calibrated and optimized during the operation for each detector phase.
Details of the detector design and its calibration can be found in \cite{Super-Kamiokande:2002weg, Abe:2013gga, Mine:2024lfh}.

\section{\label{sec:simulation}SIMULATION}
To determine the event selection criteria for the neutron decay search and evaluate the numbers of the signal and background events, Monte Carlo (MC) simulations were used.
Due to variations in the detector configuration between SK-I and SK-V, separate MC samples were generated for each period.
The propagation of particles in water is modeled using a simulation package based on GEANT3 \cite{Brun:1994aa}.
Cherenkov light production, its subsequent propagation, and the response of the PMTs and electronics are handled by a dedicated custom code.
Hadronic interactions in nuclei and water are simulated using NEUT~\cite{Hayato:2021heg} and CALOR package~\cite{Zeitnitz:1994bs}, which uses HETC~\cite{HETC}, FLUKA~\cite{FLUKA}, and MICAP~\cite{MICAP} depending on the particle and energy.

\subsection{Nucleon decay}
Neutron decay MC samples were simulated throughout the ID to account for event migration between the inside and outside of the fiducial volume.
In order to simulate the decay of a bound neutron, for instance within the oxygen nucleus of a water molecule, Fermi momentum, correlations with other nucleons, and nuclear binding energy are incorporated into the analysis.
Kaon-nucleon interactions within the oxygen nucleus and in water are also simulated in the nucleon decay MC.

The Fermi momentum distribution used in the simulation is derived from experimental data on electron scattering off $^{12}{\rm C}$ nuclei \cite{Nakamura:1976mb}.
The effective neutron mass is determined by subtracting the binding energy from the actual neutron mass.
The binding energy is modeled for each nuclear state as a Gaussian random variable with mean and standard deviation values of 39.0 (15.5)\,MeV and 10.2 (3.82)\,MeV, respectively, for the s-state (p-state) \cite{Hiramatsu:1973}.
In the nucleon decay simulation, the same binding energy is applied to both the $p_{3/2}$ and $p_{1/2}$ states in the $^{16}{\rm O}$ nucleus, which is considered a reasonable approximation given their small energy difference relative to the systematic uncertainty in the Fermi momentum.
The population ratio of neutrons in the s-state to the p-state is assumed to be 1:3, as predicted by the nuclear shell model \cite{Yamazaki:1999gz}.
Additionally, because of the correlated decay process \cite{mayer1955elementary}, neutrons in an oxygen nucleus have a 10\% probability of decaying in correlation with a neighboring nucleon.

\begin{figure}
    \centering
    \includegraphics[keepaspectratio, width=0.49\textwidth, clip]{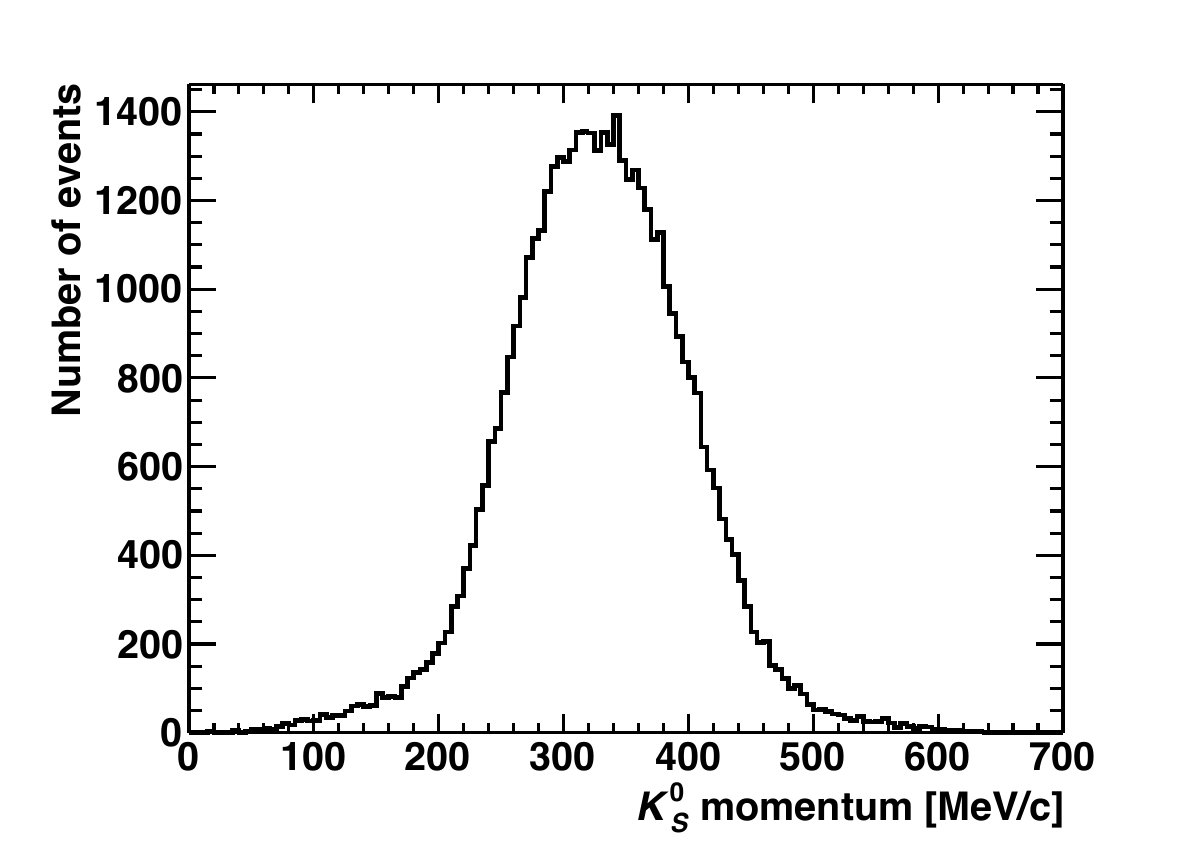}
    \caption{\label{fig:momKS}True $K^0_S$ momentum just after leaving the oxygen nucleus in the simulation.}
\end{figure}

Kaons produced from neutron decay within the nucleus may interact with nearby nucleons in the nucleus or with the water medium.
The kaon interaction model employed in this analysis has been updated for $p\to \mu^+ K^0$ channel \cite{PhysRevD.106.072003} from the previous study with SK-I \cite{PhysRevD.72.052007}.
In the nucleus, both elastic scattering and charge exchange processes of neutral kaons to nucleons are simulated.
98\% of $K^0$ from nucleon decay exit from oxygen nucleus as $K^0$.
After exiting the nucleus, propagation of a neutral kaon in water is simulated under the assumption that it propagates as a $K^0_S$ or $K^0_L$ eigenstate.
Both elastic and inelastic scattering in water are taken into account in the simulation.
As $K^0_L$ mesons with a momentum of about 300-400\,MeV/c travel approximately 10\,m in vacuum before decaying, while their mean free path in water is about 1\,m, most $K_L^0$ from neutron decay interact hadronically in water.
On the other hand, 98\% of $K^0_S$ that exit the nucleus promptly decay into $2 \pi^0$ or $\pi^+ \pi^-$ in water, with a lifetime of 90\,ps.
Figure \ref{fig:momKS} shows true $K^0_S$ momentum just after leaving the oxygen nucleus in the simulation of \nuknop.
Coherent regeneration of $K^0_L \to K^0_S$ is simulated within the oxygen nucleus and the surrounding water.
The regeneration probability is derived from kaon scattering experiments using a carbon target \cite{Eberhard:1993nb}.
The regeneration occurs in about 0.1\% of \nuknop MC events.

\subsection{Atmospheric neutrinos}
Background sources in nucleon decay search are atmospheric neutrinos.
Atmospheric neutrino MC is simulated using the atmospheric neutrino flux calculated by Honda \textit{et al}. \cite{Honda:2011nf} and NEUT 5.4.0.1 \cite{Hayato:2021heg} neutrino-nucleus event generator.
Atmospheric neutrino MC samples equivalent to 500 years of atmospheric neutrino exposure were generated to evaluate the remaining background in the final samples for each detector phase.
The main background interaction for \nuknop search is single pion production, which is simulated using the Rein-Sehgal model \cite{REIN198179}, including lepton mass correction by Berger and Sehgal \cite{Berger-Sehgal}.
The generated atmospheric neutrino events are weighted to account for the three-flavor neutrino oscillations, whose parameters are taken from \cite{ParticleDataGroup:2022pth} assuming normal mass ordering.
The updates of the kaon-nucleon interaction in nuclei and water are also included in the atmospheric neutrino MC used in this analysis.
However, the rate of neutral kaon production is less than 1\% in the remaining background events due to its small cross section and associated production, where simultaneously produced hadrons cause the event signature to fail.

\section{\label{sec:rec}Reconstruction}
The majority of triggered events are cosmic ray muons and low-energy backgrounds associated with the radioactivity of materials surrounding the detector walls.
Prior to detailed event reconstruction, multiple stages of data reduction are applied to eliminate these background events \cite{sk1_reduction}.
Reconstruction algorithms are applied to the events that remain after the reduction process.
Event reconstruction, called APFit, is employed to determine fundamental event characteristics, including the interaction vertex, number of rings, particle types, momentum, and Michel electrons using PMT hit position, integrated charge and timing information.
Cherenkov rings are categorized as either showering ($e$ or $\gamma$-like) or non-showering ($\mu$ or $\pi$-like).
Cherenkov rings produced by electrons and gammas are classified as showering due to their electromagnetic scattering and showering, resulting in diffuse rings.
In contrast, Cherenkov rings produced by muons and charged pions have sharp edges.
Those rings are identified as non-showering rings, i.e., we do not distinguish charged pions from muons.
Details of the reconstruction procedures can be found in \cite{SHIOZAWA1999240, Mine:2024lfh}.

In this analysis, we improved the reconstruction of charged pion momentum.
The performance of pion momentum reconstruction is important for the \pipi search because $K^0$ momentum and invariant mass are reconstructed from charged pion momentum and used for selection and spectrum fitting.
The electron and muon momentum are estimated from the number of observed photoelectrons within a $70^\circ$ half-opening angle around the reconstructed ring direction.
Since charged pions are more easily scattered in water than muons, their momenta are reconstructed using Cherenkov opening angle in addition to the number of observed photoelectrons.
The accuracy of the opening angle reconstruction was improved by refining the estimation of the photoelectron distribution, better accounting for the effects of hadron scattering.
Consequently, the performance of charged pion momentum reconstruction was enhanced, leading to a more precise reconstruction of the $K^0$ momentum and invariant mass.

\section{\label{sec:selection}SEARCH METHOD}
This analysis uses events with the reconstructed vertices located within the 22.5\,kton fiducial volume (more than 2\,m away from the ID wall), visible energy ($E_{\rm vis}$) above 30\,MeV, and no hit-PMT clusters in the OD, i.e., no visible energy deposit in the OD.
These events are called fully contained fiducial volume (FCFV) events.

In the \nuknop decay, the invariant mass of neutron cannot be reconstructed due to the missing momentum carried away by the antineutrino.
Instead, reconstructed $K^0$ momentum and invariant mass are used to distinguish the signal events from the atmospheric neutrino background.
In the previous search using a two-dimensional cut on these variables \cite{PhysRevD.72.052007}, more than 10 events remained in the signal region for 92\,kton$\cdot$years exposure, as expected by the atmospheric neutrino MC.
This is about two orders of magnitude larger than the number of background events expected in other decay modes with more distinctive signatures, such as $p \to e \pi^0$.
Therefore, we performed a spectral fit to the $K^0$ invariant mass distributions after signal event selections to search for excess over the background distributions in this study.
Since the invariant mass distribution has a narrower peak than the momentum distribution, the fit to the invariant mass distribution is more sensitive in this analysis.
Based on that, a spectrum fit to the $K^0$ invariant mass distributions was performed after signal event selections to search for excess over the background distributions.
Similar methods were employed for other nucleon decay searches with relatively high background levels, including $n \to \bar{\nu} \pi^0$ and $p \to \bar{\nu} \pi^+$ \cite{Super-Kamiokande:2013rwg} as well as $p \to l^+X$ ($l=e,\mu$), $n \to \nu \gamma$, and $np \to l^+ \nu$ ($l=e,\mu,\tau$) \cite{Super-Kamiokande:2015pys}.

According to the MC simulation, after the fully contained selection, 60\% of produced $K^0$ decay as $K^0_S$, 13\% decay as $K^0_L$.
The remaining 27\% do not decay as $K^0$ due to hadronic interactions that produce other particles such as $K^+$, $\Lambda$, or $\Sigma$.
Among the $K^0_L$ decay modes, the fraction of $K^0_L \to \pi^\pm l^\mp \nu$ ($l=e$ or $\mu$), which has the highest branching ratio of all $K^0_L$ decay modes, is 8\%.
However, the $K^0$ invariant mass cannot be reconstructed in this decay mode due to neutrinos carrying away energy.
Events from $K_L^0 \to 3 \pi^0$, the second most frequent decay mode, should produce up to six Cherenkov rings.
However, in most cases, no more than four Cherenkov rings can be identified.
As a result, the $K_L^0 \to 3 \pi^0$ events are actually selected in the $K_S^0 \to 2 \pi^0$ search.
Therefore, this study applies only $K^0_S$ decay selections.

There are two sets of selections in \nuknop search to extract \pizero and \pipi signals separately from the atmospheric neutrino background.
The selections are the same as in the previous analysis \cite{PhysRevD.72.052007}, except for the $K^0$ invariant mass cut which is not applied in this analysis.

\subsection{Selections for \nuknop, \pizero}
\begin{figure}
    \centering
    \includegraphics[keepaspectratio, width=0.49\textwidth, clip]{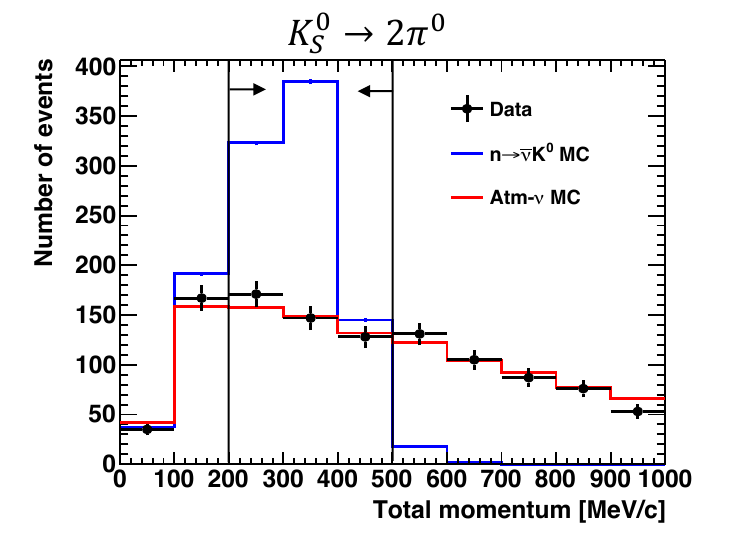}
    \includegraphics[keepaspectratio, width=0.49\textwidth, clip]{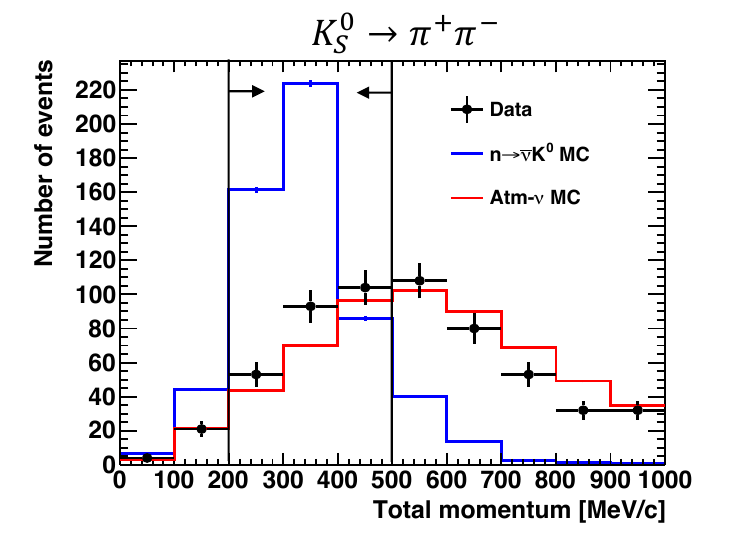}
    \caption{\label{fig:momK_sk1to5}The reconstructed $K^0$ momentum after the Michel electron selections for $K^0_S \to 2\pi^0$ (top) and $K^0_S \to \pi^+ \pi^-$ (bottom). SK-I to V are combined.
    The data (black dots) are compared with the atmospheric neutrino MC events (red) and the nucleon decay MC (blue), which are both normalized by the area of the data.
    Error bars denote statistical uncertainty.}
\end{figure}

\begin{figure*}
    \centering
    \includegraphics[keepaspectratio, width=1.0\textwidth, clip]{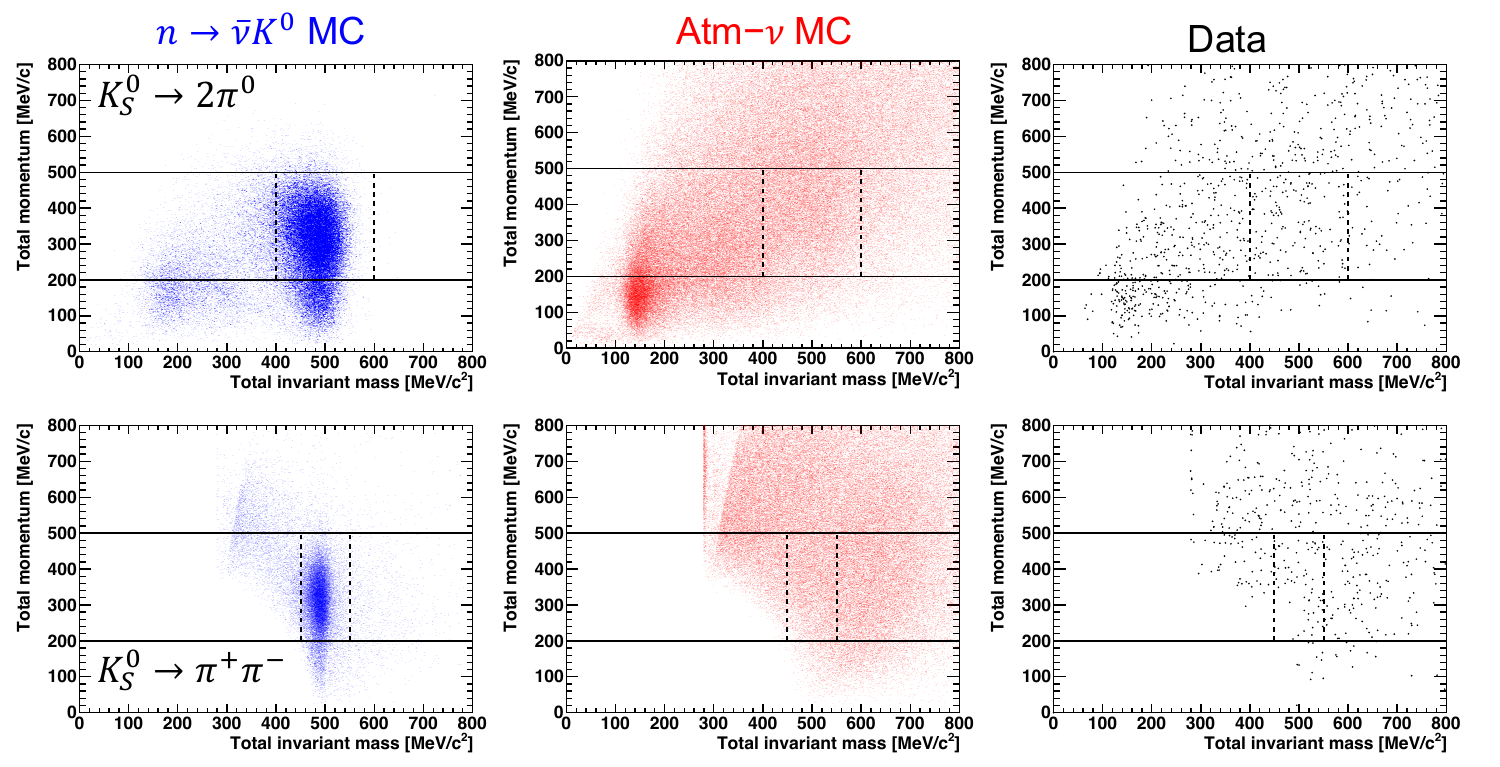}
    \caption{\label{fig:scatter_sk1to5}Scatter plots of the reconstructed $K^0$ invariant mass and $K^0$ momentum after applying the Michel electron selections for $K^0_S \to 2\pi^0$ (top) and $K^0_S \to \pi^+ \pi^-$ (bottom). SK-I to V are combined. Nucleon decay MC, atmospheric neutrino MC, and data are shown from left to right. The horizontal lines represent the cut on total momentum in this analysis. The vertical dashed lines represent the cut on total invariant mass in the previous analysis\cite{PhysRevD.72.052007}, which is not applied in this invariant mass fit analysis.}
\end{figure*}

The event selection steps for the \pizero search are applied in the following order:
\begin{enumerate}[label=A-\arabic*:]
    \item Events should be FCFV events with $E_{\rm vis}>30$\,MeV
    \item The number of rings should be three or four
    \item All rings should be showering
    \item There must be no Michel electrons
    \item The reconstructed $K^0$ momentum should be $200 < P_{K^0} < 500\,{\rm MeV/c}$
    \item The reconstructed $K^0$ invariant mass should be $300 < W_{K^0} < 700\,{\rm MeV/c^2}$
\end{enumerate}
The $K^0$ momentum distribution after applying selection A-4 is shown in the top panel of Figure \ref{fig:momK_sk1to5}.
The reconstructed $K^0$ momentum in nucleon decay MC is broadly distributed mainly due to the Fermi motion of bound neutrons.
The top panels of Figure \ref{fig:scatter_sk1to5} show scatter plots of the reconstructed $K^0$ invariant mass versus $K^0$ momentum after applying selection A-4.
The blue blob at lower invariant mass originates from \pipi decay events, which are selected as $2\pi^0$ candidate due to charge exchange from $\pi^\pm$ to $\pi^0$ and misidentification of $\pi^\pm$ as shower type.
As shown in the right panels of Figure \ref{fig:scatter_sk1to5}, a large number of data events remain in the signal region after all selections are applied.

\begin{figure}
    \centering
    \includegraphics[keepaspectratio, width=0.49\textwidth, clip]{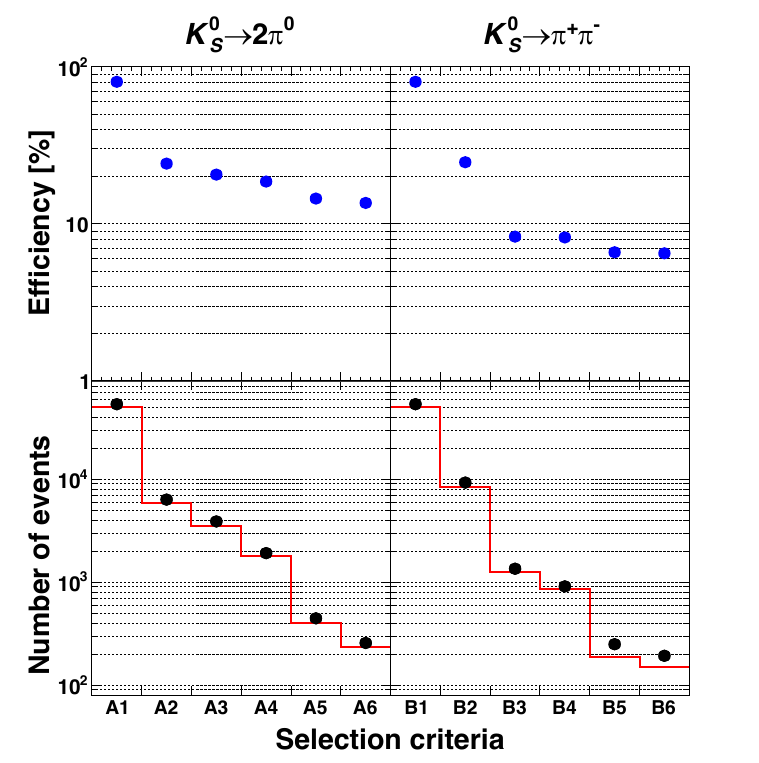}
    \caption{\label{fig:eff_bkg}Signal efficiencies (blue), number of background events (red), and number of candidates (black) for each event selection criteria. Averaged efficiency and combined number of events are shown for SK-I to V. Error bars of data points represent statistical uncertainty, but they are too small to be visible. The number of atmospheric neutrino MC events is normalized by the live time (6511.3 days).}
\end{figure}

\subsection{Selections for \nuknop, \pipi}
The event selection steps for the \pipi search are applied in the following order:
\begin{enumerate}[label=B-\arabic*:]
    \item Events should be FCFV events with $E_{\rm vis}>30$\,MeV
    \item The number of rings should be two
    \item All rings should be non-showering
    \item The number of Michel electrons should be one or zero
    \item The reconstructed $K^0$ momentum should be $200 < P_{K^0} < 500\,{\rm MeV/c}$
    \item The reconstructed $K^0$ invariant mass should be $300 < W_{K^0} < 700\,{\rm MeV/c^2}$
\end{enumerate}
The positive charged pion produced from $K^0_S \to \pi^+ \pi^-$ generates a muon from the decay, while $\pi^-$ is absorbed by oxygen nuclei in water in most cases and does not generate a muon.
Therefore, events with zero or one Michel electron are selected as signal candidates, considering the inefficiency of Michel electron tagging.
The $K^0$ momentum distribution after applying selection B-4 is shown in the bottom panel of Figure \ref{fig:momK_sk1to5}.
The slight shift between the observed data and atmospheric neutrino MC can be accounted for by the systematic uncertainties in the pion momentum in the atmospheric neutrino interactions, final state interaction (FSI) within nuclei, and secondary interaction (SI) in water, which are included in Table \ref{tab:syst_sb} and \ref{tab:syst_b}.
The bottom panels of Figure \ref{fig:scatter_sk1to5} show scatter plots of the reconstructed $K^0$ invariant mass versus $K^0$ momentum after selection B-4 is applied.
The cluster appearing with a low invariant mass around $280\,{\rm MeV/c^2}$ is due to a lower limit of the pion momentum at the Cherenkov threshold in the reconstruction algorithm.

The signal detection efficiency, expected number of backgrounds, and number of candidates are compared for each step in Figure \ref{fig:eff_bkg}.
The $K^0$ momentum cut largely reduces the number of backgrounds in both \pizero and \pipi selection.
The signal detection efficiency is defined as the ratio of the number of selected signal events to the number of generated events within the fiducial volume.
The number of expected backgrounds after B-6 selection in \pipi search is approximately 30\% lower than the number of observed data.
The excess is observed across the entire fit range, from $300\,{\rm MeV/c^2}$ to $700\,{\rm MeV/c^2}$, not being limited to the signal peak region where nucleon decay is expected.
This overall excess is resolved by considering systematic uncertainties in the spectrum fit analysis.
The difference is comparable with the systematic uncertainties, including the pion momentum reconstruction, pion interaction, and cross section for single pion production.

When the same invariant mass cuts as in the previous analysis \cite{PhysRevD.72.052007} are applied to the neutron decay MC, i.e., $400 < W_{K^0} < 600\,{\rm MeV/c^2}$ in \pizero selection and $450 < W_{K^0} < 550\,{\rm MeV/c^2}$ in \pipi selection, the signal selection efficiencies are $12.5 \pm 0.8$\% (syst.) and $5.4 \pm 1.3$\% (syst.), respectively.
The major systematic uncertainty for the efficiency of \pipi selection is the uncertainty in the pion scattering model.
Under the same selection criteria, the expected numbers of remaining background events are $123.9 \pm 28.9$ (syst.) in \pizero selection and $42.4 \pm 14.7$ (syst.) in \pipi selection.
The systematic uncertainties of the background events are suppressed to about 7\% by the sideband data in the spectrum fit described in later section.

Table \ref{tab:interaction} summarizes the interaction modes contributing to the atmospheric neutrino background after all selections are applied.
The primary background source in both \pizero and \pipi searches is single pion production.
Single eta production contributes to the background in \pizero search through the $\eta \to 3\pi^0$ decay, which has a branching ratio of 32.6\%.

\begin{table}
    \centering
    \caption{Breakdown of the interaction mode on remaining atmospheric neutrino backgrounds with MC statistical uncertainties [\%].
    SK-I to V are averaged.
    CC, QE, and DIS represent charged-current, quasi-elastic, and deep-inelastic scattering, respectively.
    In single pion production, $\pi^+$, $\pi^-$, and $\pi^0$ contributions are shown, respectively.}
    \vspace{0.2cm}
    \begin{tabular}{c c c c c} \hline \hline
	Modes & \multicolumn{2}{c}{\pizero} & \multicolumn{2}{c}{\pipi} \\ \hline
        CCQE        & \multicolumn{2}{c}{$3.2\pm0.2$} & \multicolumn{2}{c}{$16.8\pm0.6$}\\
        Single $\pi$ & \multicolumn{2}{c}{$48.6\pm0.6$} & \multicolumn{2}{c}{$64.4\pm0.7$} \\
        & $\pi^+$ & $17.1\pm0.5$ & & $46.2\pm0.8$ \\
        & $\pi^-$ & $4.6\pm0.3$  & & $13.3\pm0.5$ \\
        & $\pi^0$ & $26.9\pm0.5$ & & $4.9\pm0.3$ \\
        Single $\eta$ & \multicolumn{2}{c}{$10.2\pm0.4$} & \multicolumn{2}{c}{$0.4\pm0.1$}\\
        DIS           & \multicolumn{2}{c}{$35.5\pm0.6$} & \multicolumn{2}{c}{$14.8\pm0.5$}\\
        Others        & \multicolumn{2}{c}{$2.4\pm0.2$}  & \multicolumn{2}{c}{$3.7\pm0.3$}\\
	\hline \hline
    \end{tabular}
    \label{tab:interaction}
\end{table}

\subsection{Spectrum fit}
After event selections, a spectrum fit was performed on the reconstructed $K^0$ invariant mass distributions.
The fit minimizes the $\chi^2$, incorporating systematic uncertainties through quadratic penalty terms (pull term) \cite{PhysRevD.66.053010}.
The $\chi^2$ function is defined based on Poisson probabilities to account for the statistical fluctuation and is given by
\begin{equation}
\begin{split}    
    \label{eq:chi2}
    \chi^2 & = 2 \sum\limits_{i=1}^{N_{\rm bins}} \left[
    N^{\rm exp}_{i}
    \left( 1 + \sum_{j=1}^{N_{\rm syserr}} f_i^j \epsilon_{j} \right)
    - N_{i}^{\rm obs} \right.\\
    & + N_{i}^{\rm obs} \left.
    \ln \frac{N_{i}^{\rm obs}}{N_{i}^{\rm exp} (1 + \sum_{j=1}^{N_{\rm syserr}} f_i^j \epsilon_{j})}
    \right]
    + \sum_{j=1}^{N_{\rm syserr}} \left( \frac{\epsilon_{j}}{\sigma_{j}}\right)^2, \\
    \end{split}
\end{equation}
with,
\begin{equation}
    \label{eq:Nexp}
    N^{\rm exp}_{i} = N^{\rm bkg}_{i} + \beta N^{\rm sig}_{i},    
\end{equation}
where $i$ indexes the data bins ($i\in [1,120]$) and $j$ indexes the systematic uncertainties ($j \in [1,84]$).
For each of the five SK periods, there are 8 data bins for the $K^0$ invariant mass distribution of the \pizero samples and 16 data bins for the \pipi final samples.
The expected number of events, $N_{i}^{\rm exp}$, is obtained from the MC simulations, while $N_{i}^{\rm obs}$ represents the number of the observed events.
The MC expectation, given by Equation \ref{eq:Nexp}, consists of the expected background contribution, $N_{i}^{\rm bkg}$, and the expected signal contribution, $N_{i}^{\rm sig}$, where $\beta$ is the signal normalization factor.
In Equation \ref{eq:chi2}, systematic uncertainties are incorporated using the 84 fit error parameters $\epsilon_{j}$.
$f_i^j$ represents the fractional change in $N_{i}^{\rm exp}$ for variations corresponding to 1 sigma uncertainty $\sigma_{j}$.

\section{\label{sec:results}RESULTS}
\subsection{Systematic uncertainty}
\label{sec:syst}
Systematic uncertainties considered in this analysis are summarized in Table \ref{tab:syst_sb} and \ref{tab:syst_b}.

Physics model uncertainties in the neutron decay MC include those associated with correlated decay probabilities, Fermi momentum, pion interaction, and kaon interaction.
Uncertainties in pion FSI in nuclei and pion SI in water, considered in both the neutron decay and atmospheric neutrino MC, affect the sensitivity to a similar extent as those in the model of single pion production.
They are evaluated by changing the parameters of interaction models relevant for absorption, scattering, and charge exchange within their 1$\sigma$ constraints determined from pion scattering experiments \cite{Salcedo:1987md, Patrick_neut_pionFSI, dePerio:2014ica}.

Detector performance and reconstruction uncertainties are evaluated for both signal and background MC.
In this analysis, systematic uncertainty on the pion momentum reconstruction was revised from the previous studies.
As mentioned in Section \ref{sec:rec}, charged pion momentum is reconstructed based on the charge and opening angle of the Cherenkov ring.
For the charge measurement, energy scale uncertainties of the electron and muon momentum have been evaluated from a comparison of the absolute energy scales between control sample data and MC, which include cosmic-ray muons
, neutral pions from atmospheric neutrino interactions, and Michel electrons from cosmic-ray muons.
These uncertainties are in the range of about 2--3\% for each SK period.
On the other hand, to assess the uncertainty in the estimation of the opening angle, we compared pion momentum reconstructed using only the opening angle in multi $\mu$-like ring samples with two Michel electrons from atmospheric neutrino data and MC.
These samples are independent of the signal events in \nuknop search, which requires at most one Michel electron.
The total uncertainty in pion momentum reconstruction is derived by combining these contributions from the estimations by the charge and opening angle in quadrature.
Pion momentum uncertainties vary by detector periods, from 2.9\% in SK-IV to 8.1\% in SK-III, depending on the statistics of multi $\mu$-like ring sample.

Uncertainties related to atmospheric neutrino flux, interactions, and oscillations are accounted for based on nearly the same sources as in the atmospheric neutrino analysis \cite{Super-Kamiokande:2023ahc}, with detailed descriptions available in \cite{Wester:2023kac}.
Among these, systematic uncertainties in the single pion production model have the largest impact on the sensitivity.

\begin{figure}
    \centering
    \includegraphics[keepaspectratio, width=0.49\textwidth, clip]{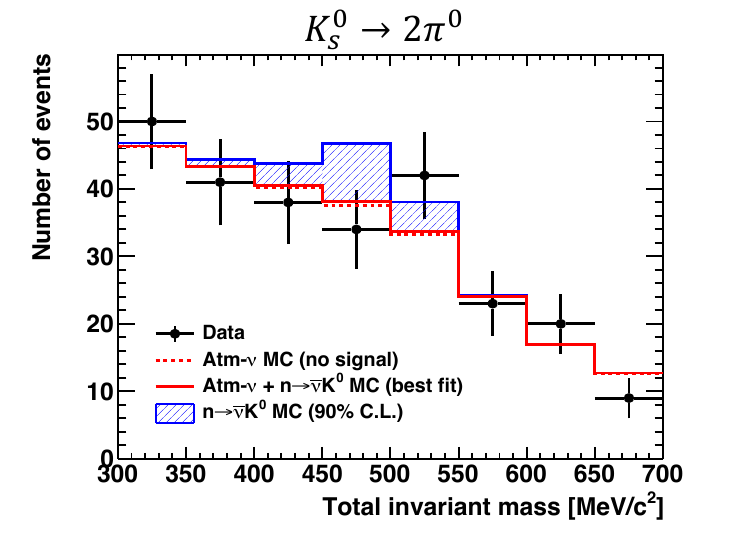}
    \includegraphics[keepaspectratio, width=0.49\textwidth, clip]{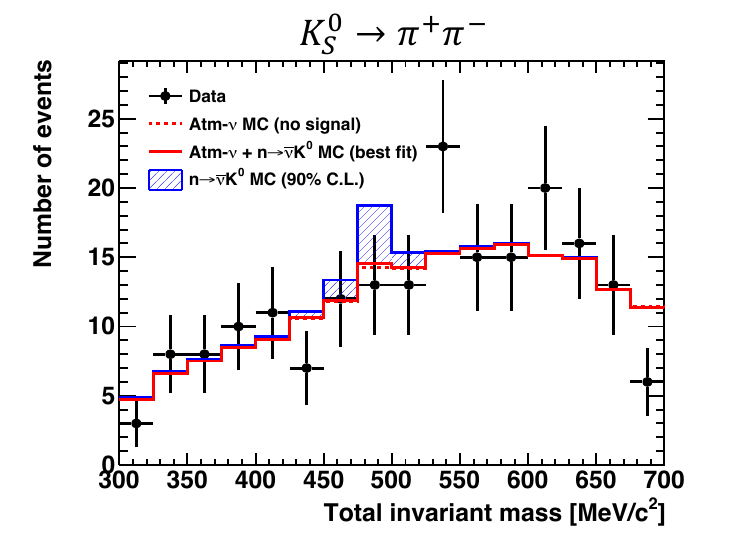}
    \caption{\label{fig:massK_sk1to5_afterfit}The reconstructed $K^0$ invariant mass distributions with the spectrum fit results for $K^0_S \to 2\pi^0$ (top) and $K^0_S \to \pi^+ \pi^-$ (bottom). SK-I to V are combined. The data (black dots) are compared with the atmospheric neutrino MC under the assumption of no signal (dashed red line) and the atmospheric neutrino plus nucleon decay MC simulation at best-fit (solid red line). The 90\% C.L. upper limit of the nucleon decay events (blue hatched histogram) is also displayed.}
\end{figure}

\subsection{Fitting Results}
\label{sec:fitresults}
The minimization of $\chi^2$ was performed with respect to each $\beta$ (the signal normalization factor) by solving $\partial \chi^2 / \partial \epsilon_{j} = 0$ in Equation \ref{eq:chi2}.

The best-fit $\chi^2 / N_{\rm dof}$ value was $117.4/119$ with a value of $\beta$ corresponding to a total of 1.7 signal events in 0.401 Mton$\cdot$years exposure.
It is also consistent with the expectation of 0 signal events, and there is no significant excess of signals over the background.
As shown in Table \ref{tab:syst_sb} and \ref{tab:syst_b}, there are no systematic error parameters that show a significantly larger deviation than the estimated uncertainty.
Using the fitted values and pull terms, the combined invariant mass distributions for SK-I to SK-V are shown in Figure \ref{fig:massK_sk1to5_afterfit}.
The data are compared to the atmospheric neutrino plus nucleon decay MC simulation with corrections by the best-fit parameters.
The 90\% C.L. upper limit of the nucleon decay signal events is also shown with blue hatched histograms.
The upper limit on the number of signal events is 27.7 in total at 90\% C.L.
Since charged pions have better momentum resolution than $\gamma$'s from the decay of neutral pions, the \pipi mass distribution (Figure \ref{fig:massK_sk1to5_afterfit}, bottom) has a better mass resolution than that of \pizero (Figure \ref{fig:massK_sk1to5_afterfit}, top), exhibiting a narrower peak around $497.6\,{\rm MeV/c^2}$.
The momentum resolution is about 3\% for charged pions with typical momenta of 200-400\,MeV/c in events that passed the B-4 selection, whereas it is about 12\% for $\gamma$'s with typical momenta of 0-300\,MeV/c in events that passed the A-4 selection.
To reflect this improved resolution, the bin width for the \pipi distribution is set to half that of the \pizero distribution.

As no significant signals were found, we set a lower limit on the lifetime of a bound neutron.
The lower lifetime limit is calculated using
\begin{equation}
    \tau /\mathcal{B}  = \frac{\Delta t N_{\rm neutrons}}{\sum_{i= \rm SK1}^{\rm SK5} N_{90 \rm CL}^{i} / \eta_{i}}, 
\end{equation}
where $\mathcal{B}$ is the branching ratio of the nucleon decay mode, $i$ indexes the SK detector phases, $\Delta t$ is the total exposure in kton$\cdot$years, and $N_{\rm neutrons}$ represents the number of neutrons per kiloton of water ($2.68 \times 10^{32}\, {\rm neutrons/kton}$).
$N_{90 \rm CL}^{i}$ is the upper limit of the number of nucleon decay events ($K^0_S \to 2\pi^0$ and $K^0_S \to \pi^+ \pi^-$) in each SK period at 90\% C.L., determined from the fit, and $\eta_{i}$ is the total signal efficiency for the two decay modes in each SK period.
We set a new lower lifetime limit for $n \to \bar{\nu} + K^0$ as $\tau/\mathcal{B} > 7.8 \times 10^{32}$\,years at 90\% C.L.
This limit is six times more stringent than our previous published limit of $\tau/\mathcal{B} > 1.3 \times 10^{32}$\,years\cite{PhysRevD.72.052007}.

\section{\label{sec:conclusion}CONCLUSION}
A search for neutron decay into an antineutrino and a neutral kaon was conducted using data from the Super-Kamiokande experiment.
No statistically significant signal excess has been observed in the signal region using all available data from pure water phases.
By performing a spectrum fit to the $K^0$ invariant mass distributions, we set a lower lifetime limit of $7.8\times10^{32}$\,years at 90\% C.L.
This result improves the previous limit by a factor of six, making it the most stringent constraint on the \nuk decay mode to date.

\begin{acknowledgments}
\input{SK-paper-acknowledgements-20250530}

\end{acknowledgments}

\appendix*

\section{SYSTEMATIC UNCERTAINTIES}

\begin{table*}
    \centering
    \caption{Systematic uncertainties related to physics model and reconstruction for both nucleon decay and atmospheric neutrino MC.
    The uncertainties on correlated decay, Fermi momentum, and kaon interaction are considered in only nucleon decay MC.
    The best-fit pull value of each systematic uncertainty parameter and the estimated 1$\sigma$ uncertainty are shown.}
    \vspace{0.2cm}
    \begin{tabular}{l c c c c c c c c c c} \hline \hline
        Systematic uncertainty & \multicolumn{5}{c}{Fit pull value ($\sigma$)} & \multicolumn{5}{c}{1$\sigma$ uncertainty (\%)} \\ \hline
        Correlated decay & \multicolumn{5}{c}{0.028} & \multicolumn{5}{c}{100} \\
        Fermi momentum & \multicolumn{5}{c}{0.0046} & \multicolumn{5}{c}{10} \\
        Pion interaction for \pizero & \multicolumn{5}{c}{-0.15} & \multicolumn{5}{c}{10} \\
        Pion interaction for \pipi & \multicolumn{5}{c}{-0.53} & \multicolumn{5}{c}{10} \\
        Kaon interaction (FSI) & \multicolumn{5}{c}{0.030} & \multicolumn{5}{c}{25} \\
        Kaon interaction (SI) & \multicolumn{5}{c}{0.031} & \multicolumn{5}{c}{50} \\
        Ring counting & \multicolumn{5}{c}{0.035} & \multicolumn{5}{c}{10} \\
        Particle identification & \multicolumn{5}{c}{0.14} & \multicolumn{5}{c}{10} \\
        Michel electron tagging & \multicolumn{5}{c}{0.17} & \multicolumn{5}{c}{10} \\ \hline
         & SK-I & SK-II & SK-III & SK-IV & SK-V & SK-I & SK-II & SK-III & SK-IV & SK-V \\ \hline
        FC reduction & 0.0003 & 0.0000 & 0.0007 & 0.0000 & -0.0003 & 0.2 & 0.2 & 0.8 & 1.3 & 1.7 \\
        Fiducial volume & 0.013 & 0.08 & 0.20 & -0.25 & -0.0051 & 2 & 2 & 2 & 2 & 2 \\
        Non-$\nu_e$ background & 0.0002 & 0.0000 & 0.0004 & 0.0000 & -0.0001 & 1 & 1 & 1 & 1 & 1 \\
        Energy scale & 0.24 & 0.23 & 0.13 & -0.042 & -0.019 & 3.3 & 2 & 2.4 & 2.1 & 1.8 \\
        Up/down energy calibration & 0.047 & -0.23 & -0.42 & 0.02 & 0.59 & 0.6 & 1.1 & 0.6 & 0.5 & 0.69 \\
        Pion momentum reconstruction & -0.70 & -0.33 & -1.0 & 0.37 & 0.37 & 5.9 & 4.6 & 8.1 & 2.9 & 5.1 \\
	\hline \hline
    \end{tabular}
    \label{tab:syst_sb}
\end{table*}

\begin{table*}
    \centering
    \caption{Systematic uncertainties related to neutrino flux, interaction, and oscillation for atmospheric neutrino MC only. The best-fit pull value of each systematic uncertainty and the estimated 1$\sigma$ uncertainty are shown.}
    \vspace{0.2cm}
    \begin{tabular}{l l c c} \hline \hline
        Systematic uncertainty &  & Fit pull value ($\sigma$) & 1$\sigma$ uncertainty (\%) \\ \hline
        Flux normalization & $E_\nu < 1\,{\rm GeV}$ & 0.15 & 7--25\footnote[1]{Uncertainty decreases linearly with $\log E_{\nu}$ from 25\% at 0.1\,GeV to 7\% at 1\,GeV.} \\
         & $E_\nu > 1\,{\rm GeV}$ & 0.12 & 7--20\footnote[2]{Uncertainty is 7\% up to 10\,GeV, increases linearly with $\log E_{\nu}$ from 7\% at 10\,GeV to 12\% at 100\,GeV, and then to 20\% at 1\,TeV.} \\
        $(\nu_\mu + \bar{\nu}_\mu)/(\nu_e + \bar{\nu}_e)$ ratio & $E_\nu < 1\,{\rm GeV}$ & 0.012 & 2 \\
         & $1 < E_\nu < 10\,{\rm GeV}$ & 0.008 & 3 \\
         & $E_\nu > 10\,{\rm GeV}$ & 0.000 & 5--30\footnote[3]{Uncertainty increases linearly with $\log E_{\nu}$ from 5\% at 30\,GeV to 30\% at 1\,TeV.} \\
        $\nu_e/\bar{\nu}_e$ ratio & $E_\nu < 1\,{\rm GeV}$ & 0.006 & 5 \\
         & $1 < E_\nu < 10\,{\rm GeV}$ & 0.018 & 5 \\
         & $E_\nu > 10\,{\rm GeV}$ & 0.003 & 8--20\footnote[4]{Uncertainty increases linearly with $\log E_{\nu}$ from 8\% at 100\,GeV to 20\% at 1\,TeV.} \\
        $\nu_\mu/\bar{\nu}_\mu$ ratio & $E_\nu < 1\,{\rm GeV}$ & 0.009 & 2 \\
         & $1 < E_\nu < 10\,{\rm GeV}$ & 0.011 & 6 \\
         & $E_\nu > 10\,{\rm GeV}$ & 0.002 & 6--40\footnote[5]{Uncertainty increases linearly with $\log E_{\nu}$ from 6\% at 50\,GeV to 40\% at 1\,TeV.} \\
        Up/down ratio &  & 0.000 & 1 \\
        Horizontal/vertical ratio &  & 0.000 & 1 \\
        $K/\pi$ ratio &  & 0.012 & 5--20\footnote[6]{Uncertainty is 5\% up to 100\,GeV, increases linearly with $\log E_{\nu}$ from 5\% at 100\,GeV to 20\% at 1\,TeV.} \\
        Neutrino path length &  & 0.009 & 10 \\
        Matter effect &  & -0.003 & 6.8 \\
        Solar activity of SK-I, II, III, IV, V &  & 0.005, 0.040, 0.025, -0.021, -0.001 & 20, 50, 20, 7, 20 \\
        CCQE & $M^{QE}_{A}$ & -0.18 & 10 \\
        & Cross section shape & 0.026 & 10 \\
        & Normalization (sub-GeV) & 0.002 & 10 \\
        & Normalization (multi-GeV) & -0.15 & 10 \\
         & $\nu/\bar{\nu}$ ratio & 0.006 & 10 \\
         & $\nu_\mu/\nu_e$ ratio & -0.022 & 10 \\
         & MEC on/off & -0.069 & 10 \\
        Single pion production & $\pi^0/\pi^+$ ratio & -0.20 & 40 \\
         & $\nu/\bar{\nu}$ ratio & -0.096 & 10 \\
         & Axial coupling & 0.35 & 10 \\
         & $C^5_A$ & 0.28 & 10 \\
         & Background & 0.075 & 10 \\
         & Coherent $\pi$ & 0.004 & 100 \\
        DIS & Model difference & 0.28 & 10 \\
         & Cross section & 0.039 & 10 \\
         & $Q^2$ distribution (high $W$) & 0.036 & 10 \\
         & $Q^2$ distribution (low $W$) & 0.010 & 10 \\
         & $Q^2$ distribution (vector) & -0.028 & 10 \\
         & $Q^2$ distribution (axial) & -0.009 & 10 \\
         & Hadron multiplicity & 0.092 & 10 \\
        Other cross section & NC/CC ratio & 0.44 & 20 \\
        Oscillation & $\Delta{}m^2_{21}$ & -0.003 & 0.00018 \\
         & $\sin^2\theta_{12}$ & -0.002 & 1.3 \\
         & $\sin^2\theta_{13}$ & -0.001 & 0.07 \\
	\hline \hline
    \end{tabular}
    \label{tab:syst_b}
\end{table*}

\nocite{}
\bibliography{nuK0_main}

\end{document}

%% file: authors-20250610.tex
\newcommand{\AFFicrr}{\affiliation{Kamioka Observatory, Institute for Cosmic Ray Research, University of Tokyo, Kamioka, Gifu 506-1205, Japan}}
\newcommand{\AFFkashiwa}{\affiliation{Research Center for Cosmic Neutrinos, Institute for Cosmic Ray Research, University of Tokyo, Kashiwa, Chiba 277-8582, Japan}}
\newcommand{\AFFipmu}{\affiliation{Kavli Institute for the Physics and
Mathematics of the Universe (WPI), The University of Tokyo Institutes for Advanced Study,
University of Tokyo, Kashiwa, Chiba 277-8583, Japan }}
\newcommand{\AFFmad}{\affiliation{Department of Theoretical Physics, University Autonoma Madrid, 28049 Madrid, Spain}}
\newcommand{\AFFubc}{\affiliation{Department of Physics and Astronomy, University of British Columbia, Vancouver, BC, V6T1Z4, Canada}}
\newcommand{\AFFbu}{\affiliation{Department of Physics, Boston University, Boston, MA 02215, USA}}
\newcommand{\AFFuci}{\affiliation{Department of Physics and Astronomy, University of California, Irvine, Irvine, CA 92697-4575, USA }}
\newcommand{\AFFcsu}{\affiliation{Department of Physics, California State University, Dominguez Hills, Carson, CA 90747, USA}}
\newcommand{\AFFcnm}{\affiliation{Institute for Universe and Elementary Particles, Chonnam National University, Gwangju 61186, Korea}}
\newcommand{\AFFduke}{\affiliation{Department of Physics, Duke University, Durham NC 27708, USA}}
\newcommand{\AFFgifu}{\affiliation{Department of Physics, Gifu University, Gifu, Gifu 501-1193, Japan}}
\newcommand{\AFFgist}{\affiliation{GIST College, Gwangju Institute of Science and Technology, Gwangju 500-712, Korea}}
\newcommand{\AFFuh}{\affiliation{Department of Physics and Astronomy, University of Hawaii, Honolulu, HI 96822, USA}}
\newcommand{\AFFicl}{\affiliation{Department of Physics, Imperial College London , London, SW7 2AZ, United Kingdom }}
\newcommand{\AFFkek}{\affiliation{High Energy Accelerator Research Organization (KEK), Tsukuba, Ibaraki 305-0801, Japan }}
\newcommand{\AFFkobe}{\affiliation{Department of Physics, Kobe University, Kobe, Hyogo 657-8501, Japan}}
\newcommand{\AFFkyoto}{\affiliation{Department of Physics, Kyoto University, Kyoto, Kyoto 606-8502, Japan}}
\newcommand{\AFFliv}{\affiliation{Department of Physics, University of Liverpool, Liverpool, L69 7ZE, United Kingdom}}
\newcommand{\AFFmiyagi}{\affiliation{Department of Physics, Miyagi University of Education, Sendai, Miyagi 980-0845, Japan}}
\newcommand{\AFFnagoya}{\affiliation{Institute for Space-Earth Environmental Research, Nagoya University, Nagoya, Aichi 464-8602, Japan}}
\newcommand{\AFFkmi}{\affiliation{Kobayashi-Maskawa Institute for the Origin of Particles and the Universe, Nagoya University, Nagoya, Aichi 464-8602, Japan}}
\newcommand{\AFFpol}{\affiliation{National Centre For Nuclear Research, 02-093 Warsaw, Poland}}
\newcommand{\AFFsuny}{\affiliation{Department of Physics and Astronomy, State University of New York at Stony Brook, NY 11794-3800, USA}}
\newcommand{\AFFokayama}{\affiliation{Department of Physics, Okayama University, Okayama, Okayama 700-8530, Japan }}
\newcommand{\AFFosaka}{\affiliation{Department of Physics, Osaka University, Toyonaka, Osaka 560-0043, Japan}}
\newcommand{\AFFox}{\affiliation{Department of Physics, Oxford University, Oxford, OX1 3PU, United Kingdom}}
\newcommand{\AFFqmul}{\affiliation{School of Physics and Astronomy, Queen Mary University of London, London, E1 4NS, United Kingdom}}
\newcommand{\AFFregina}{\affiliation{Department of Physics, University of Regina, 3737 Wascana Parkway, Regina, SK, S4SOA2, Canada}}
\newcommand{\AFFseoul}{\affiliation{Department of Physics and Astronomy, Seoul National University, Seoul 151-742, Korea}}
\newcommand{\AFFsheff}{\affiliation{School of Mathematical and Physical Sciences, University of Sheffield, S3 7RH, Sheffield, United Kingdom}}
\newcommand{\AFFshizuokasc}{\affiliation{Department of Informatics in
Social Welfare, Shizuoka University of Welfare, Yaizu, Shizuoka, 425-8611, Japan}}
\newcommand{\AFFstfc}{\affiliation{STFC, Rutherford Appleton Laboratory, Harwell Oxford, and Daresbury Laboratory, Warrington, OX11 0QX, United Kingdom}}
\newcommand{\AFFskk}{\affiliation{Department of Physics, Sungkyunkwan University, Suwon 440-746, Korea}}
\newcommand{\AFFtodai}{\affiliation{Department of Physics, University of Tokyo, Bunkyo, Tokyo 113-0033, Japan }}
\newcommand{\AFFtit}{\affiliation{Department of Physics, Institute of Science Tokyo, Meguro, Tokyo 152-8551, Japan }}
\newcommand{\AFFtus}{\affiliation{Department of Physics and Astronomy, Faculty of Science and Technology, Tokyo University of Science, Noda, Chiba 278-8510, Japan }}
\newcommand{\AFFtriumf}{\affiliation{TRIUMF, 4004 Wesbrook Mall, Vancouver, BC, V6T2A3, Canada }}
\newcommand{\AFFtokai}{\affiliation{Department of Physics, Tokai University, Hiratsuka, Kanagawa 259-1292, Japan}}
\newcommand{\AFFtsinghua}{\affiliation{Department of Engineering Physics, Tsinghua University, Beijing, 100084, China}}
\newcommand{\AFFynu}{\affiliation{Department of Physics, Yokohama National University, Yokohama, Kanagawa, 240-8501, Japan}}
\newcommand{\AFFllr}{\affiliation{Ecole Polytechnique, IN2P3-CNRS, Laboratoire Leprince-Ringuet, F-91120 Palaiseau, France }}
\newcommand{\AFFbari}{\affiliation{ Dipartimento Interuniversitario di Fisica, INFN Sezione di Bari and Universit\`a e Politecnico di Bari, I-70125, Bari, Italy}}
\newcommand{\AFFnapoli}{\affiliation{Dipartimento di Fisica, INFN Sezione di Napoli and Universit\`a di Napoli, I-80126, Napoli, Italy}}
\newcommand{\AFFroma}{\affiliation{INFN Sezione di Roma and Universit\`a di Roma ``La Sapienza'', I-00185, Roma, Italy}}
\newcommand{\AFFpadova}{\affiliation{Dipartimento di Fisica, INFN Sezione di Padova and Universit\`a di Padova, I-35131, Padova, Italy}}
\newcommand{\AFFkeio}{\affiliation{Department of Physics, Keio University, Yokohama, Kanagawa, 223-8522, Japan}}
\newcommand{\AFFwinnipeg}{\affiliation{Department of Physics, University of Winnipeg, MB R3J 3L8, Canada }}
\newcommand{\AFFkcl}{\affiliation{Department of Physics, King's College London, London, WC2R 2LS, UK }}
\newcommand{\AFFwarwick}{\affiliation{Department of Physics, University of Warwick, Coventry, CV4 7AL, UK }}
\newcommand{\AFFral}{\affiliation{Rutherford Appleton Laboratory, Harwell, Oxford, OX11 0QX, UK }}
\newcommand{\AFFwu}{\affiliation{Faculty of Physics, University of Warsaw, Warsaw, 02-093, Poland }}
\newcommand{\AFFbcit}{\affiliation{Department of Physics, British Columbia Institute of Technology, Burnaby, BC, V5G 3H2, Canada }}
\newcommand{\AFFtohoku}{\affiliation{Department of Physics, Faculty of Science, Tohoku University, Sendai, Miyagi, 980-8578, Japan }}
\newcommand{\AFFicise}{\affiliation{Institute For Interdisciplinary Research in Science and Education, ICISE, Quy Nhon, 55121, Vietnam }}
\newcommand{\AFFilance}{\affiliation{ILANCE, CNRS - University of Tokyo International Research Laboratory, Kashiwa, Chiba 277-8582, Japan}}
\newcommand{\AFFibs}{\affiliation{Center for Underground Physics, Institute for Basic Science (IBS), Daejeon, 34126, Korea}}
\newcommand{\AFFglasgow}{\affiliation{School of Physics and Astronomy, University of Glasgow, Glasgow, Scotland, G12 8QQ, United Kingdom}}
\newcommand{\AFFoecu}{\affiliation{Media Communication Center, Osaka Electro-Communication University, Neyagawa, Osaka, 572-8530, Japan}}
\newcommand{\AFFminn}{\affiliation{School of Physics and Astronomy, University of Minnesota, Minneapolis, MN  55455, USA}}
\newcommand{\AFFsilesia}{\affiliation{August Che\l{}kowski Institute of Physics, University of Silesia in Katowice, 75 Pu\l{}ku Piechoty 1, 41-500 Chorz\'{o}w, Poland}}
\newcommand{\AFFtoyama}{\affiliation{Faculty of Science, University of Toyama, Toyama City, Toyama 930-8555, Japan}}
\newcommand{\AFFbmcc}{\affiliation{Science Department, Borough of Manhattan Community College / City University of New York, New York, New York, 1007, USA.}}
\newcommand{\AFFnumazu}{\affiliation{National Institute of Technology, Numazu College, Numazu, Shizuoka 410-8501, Japan}}

\AFFicrr
\AFFkashiwa
\AFFmad
\AFFbmcc
\AFFbu
\AFFbcit
\AFFuci
\AFFcsu
\AFFcnm
\AFFduke
\AFFllr
\AFFgifu
\AFFgist
\AFFglasgow
\AFFuh
\AFFibs
\AFFicise
\AFFicl
\AFFbari
\AFFnapoli
\AFFpadova
\AFFroma
\AFFilance
\AFFkeio
\AFFkek
\AFFkcl
\AFFkobe
\AFFkyoto
\AFFliv
\AFFminn
\AFFmiyagi
\AFFnagoya
\AFFkmi
\AFFpol
\AFFnumazu
\AFFsuny
\AFFokayama
\AFFoecu
\AFFox
\AFFral
\AFFseoul
\AFFsheff
\AFFshizuokasc
\AFFsilesia
\AFFstfc
\AFFskk
\AFFtohoku
\AFFtodai
\AFFipmu
\AFFtit
\AFFtus
\AFFtoyama
\AFFtriumf
\AFFtsinghua
\AFFwu
\AFFwarwick
\AFFwinnipeg
\AFFynu

\author{K.~Yamauchi\,\orcidlink{0009-0000-0112-0619}}
\AFFtus

\author{K.~Abe}
\AFFicrr
\AFFipmu
\author{S.~Abe}
\AFFicrr
\author{Y.~Asaoka}
\AFFicrr
\AFFipmu
\author{M.~Harada}
\AFFicrr
\author{Y.~Hayato}
\AFFicrr
\AFFipmu
\author{K.~Hiraide}
\AFFicrr
\AFFipmu
\author{K.~Hosokawa}
\AFFicrr
\author{K.~Ieki}
\author{M.~Ikeda}
\AFFicrr
\AFFipmu
% ...


\collaboration{The Super-Kamiokande Collaboration}
\noaffiliation

%% file: SK-paper-acknowledgements-20250530.tex


We gratefully acknowledge the cooperation of the Kamioka Mining and Smelting Company. The Super-Kamiokande experiment has been built and operated from funding by the Japanese Ministry of Education, Culture, Sports, Science and Technology; the U.S. Department of Energy; and the U.S. National Science Foundation. Some of us have been supported by funds from the National Research Foundation of Korea (NRF-2009-0083526, NRF-2022R1A5A1030700, NRF-2202R1A3B1078756, RS-2025-00514948) funded by the Ministry of Science, Information and Communication Technology (ICT); the Institute for Basic Science (IBS-R016-Y2); and the Ministry of Education (2018R1D1A1B07049158, 2021R1I1A1A01042256, RS-2024-00442775); the Japan Society for the Promotion of Science; the National Natural Science Foundation of China under Grants No. 12375100; the Spanish Ministry of Science, Universities and Innovation (grant PID2021-124050NB-C31); the Natural Sciences and Engineering Research Council (NSERC) of Canada; the Scinet and Digital Research of Alliance Canada; the National Science Centre (UMO-2018/30/E/ST2/00441 and UMO-2022/46/E/ST2/00336) and the Ministry of  Science and Higher Education (2023/WK/04), Poland; the Science and Technology Facilities Council (STFC) and Grid for Particle Physics (GridPP), UK; the European Union’s Horizon 2020 Research and Innovation Programme H2020-MSCA-RISE-2018 JENNIFER2 grant agreement no.822070, H2020-MSCA-RISE-2019 SK2HK grant agreement no. 872549; and European Union's Next Generation EU/PRTR  grant CA3/RSUE2021-00559; the National Institute for Nuclear Physics (INFN), Italy.